\documentclass[12pt]{article}
\usepackage{graphicx}

\newcommand{\beq}{\begin{equation}}
\newcommand{\eeq}{\end{equation}}
\newcommand{\beqa}{\begin{eqnarray}}
\newcommand{\eeqa}{\end{eqnarray}}
\newcommand{\non}{\nonumber}
\newcommand{\lab}{\label}

\newcommand{\ket}{\rangle}

\begin{document}

\title{Interaction-free quantum computation}

\author{Hiroo Azuma\thanks{On leave from
Centre for Quantum Computation,
Clarendon Laboratory,
Parks Road, Oxford OX1 3PU, United Kingdom.}
\\
{\small Canon Inc., 5-1, Morinosato-Wakamiya,}\\
{\small Atsugi-shi, Kanagawa, 243-0193, Japan}\\
{\small E-mail: azuma.hiroo@canon.co.jp}}

\date{March 23, 2004}

\maketitle

\begin{abstract}
In this paper, we study the quantum computation realized
by an interaction-free measurement (IFM).
Using Kwiat et al.'s interferometer,
we construct a two-qubit quantum gate
that changes one particle's trajectory
according to whether or not the other particle exists
in the interferometer.
We propose a method for distinguishing Bell-basis vectors,
each of which consists of a pair of an electron and 
a positron, by this gate.
(This is called the Bell-basis measurement.)
This method succeeds with probability $1$
in the limit of $N \rightarrow \infty$,
where $N$ is the number of beam splitters
in the interferometer.
Moreover, we can carry out a controlled-NOT gate
operation by the above Bell-basis measurement and
the method proposed by Gottesman and Chuang.
Therefore, we can prepare a universal set of
quantum gates by the IFM.
This means that we can execute any quantum algorithm
by the IFM.
\end{abstract}

\section{Introduction}
\lab{introduction}
Since many researchers recognized
that a quantum computer solves certain problems
more efficiently than any conventional computer,
quantum computation has been studied eagerly
\cite{quantum-algorithms}.
The quantum computation is process of successive unitary
transformations and measurements
to two-state systems that are called qubits.
Any unitary transformation applied to qubits can be
decomposed into $U(2)$ transformations for one qubit
and controlled-NOT (CNOT) gates for two qubits
\cite{Barenco-etal}.
The CNOT gate generates entanglement between two qubits.
Many researchers propose various methods for realizing
the CNOT gate and carry out experiments
\cite{CNOT-gate}.

The Bell-basis measurement, which distinguishes
Bell-basis vectors $\{|\Phi^{\pm}\ket,|\Psi^{\pm}\ket\}$,
is a basic operation in quantum information processing.
It plays an important role in the quantum teleportation
\cite{QuantumTeleportation}.
Gottesman and Chuang showed
that we can carry out the CNOT gate
if we prepare a certain special four-qubit state $|\chi\ket$
and execute the Bell-basis measurement twice
\cite{Gottesman-Chuang}.
This implies the following fact.
Let us assume that it is difficult to carry out the CNOT
gate directly in a certain system.
Even in this system,
if we can accomplish the Bell-basis measurement easily,
we can carry out the CNOT gate indirectly
through Gottesman and Chuang's method.

In this paper, we consider the Bell-basis measurement
realized by an interaction-free measurement (IFM).
Using the IFM, we can construct a two-qubit gate
that changes one particle's trajectory
according to whether or not the other particle exists
in the interferometer.
By this two-qubit gate, we can generate the Bell states
and a certain four-qubit state $|\chi\ket$
introduced by Gottesman and Chuang
\cite{Kwiat-etal-1996,Azuma}.
Hence, if we can carry out the Bell-basis measurement
by the IFM,
we can accomplish the CNOT gate indirectly.
This is motivation of this paper.

This paper is organized as follows.
In the latter half of this section,
we review the IFM briefly.
In Sec.~\ref{bell-basis-measurement-IFM},
we consider the Bell-basis measurement by the IFM process.
In Sec.~\ref{cnot-gate-IFM},
we discuss implementation of the CNOT gate by the IFM process.
In Sec.~\ref{discussion},
we give a brief discussion.

Here, we review the IFM.
The IFM was proposed by Dicke, Elitzur, and Vaidman first
\cite{Dicke,Elitzur-Vaidman}.
Then, it is refined by Kwiat et al.
\cite{Kwiat-etal-IFM}.
The IFM is a method for examining whether or not
an object exists in an interferometer.
We assume that the object absorbs a photon
when the photon approaches the object closely enough.
We have to detect the object without its absorption.

Kwiat et al. consider an interferometer that consists
of $N$ beam splitters as shown in Fig.~\ref{KWinterferometer3}.
We describe the upper paths as $a$
and lower paths as $b$,
so that the beam splitters form the boundary line
between the paths $a$ and the paths $b$
in the interferometer.
We write a state with one photon on the paths $a$
as $|1\ket_{a}$,
and a state with no photon on the paths $a$ as $|0\ket_{a}$.
This notation applies to the paths $b$ as well.
The beam splitter $B$ in Fig.~\ref{KWinterferometer3}
works as follows:
\beq
B:
\left\{
\begin{array}{rrr}
|1\ket_{a}|0\ket_{b} & \rightarrow &
\cos\theta|1\ket_{a}|0\ket_{b}-\sin\theta|0\ket_{a}|1\ket_{b}\\
|0\ket_{a}|1\ket_{b} & \rightarrow &
\sin\theta|1\ket_{a}|0\ket_{b}+\cos\theta|0\ket_{a}|1\ket_{b}
\end{array}
\right..
\lab{definition-beam-splitter-B}
\eeq
(The transmissivity of $B$ is given by $T=\sin^{2}\theta$,
and the reflectivity of $B$ is given by $R=\cos^{2}\theta$
in Eq.~(\ref{definition-beam-splitter-B}).)

\begin{figure}
\begin{center}
\includegraphics[scale=0.9]{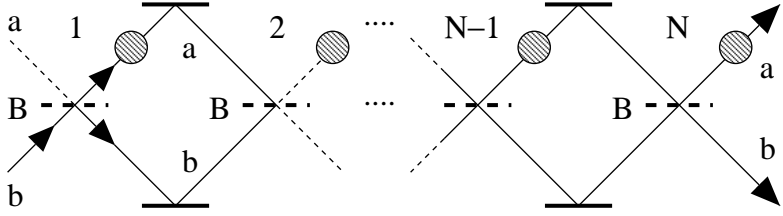}
\end{center}
\caption{Kwiat et al.'s interferometer for the IFM.}
\lab{KWinterferometer3}
\end{figure}

Let us throw a photon into the lower left port of $b$
in Fig.~\ref{KWinterferometer3}.
If there is no object on the paths,
the wave function of the photon that comes from the $k$th
beam splitter is given by
\beq
\sin k\theta|1\ket_{a}|0\ket_{b}
+\cos k\theta|0\ket_{a}|1\ket_{b}
\quad\quad
\mbox{for $k=0,1,...,N$.}
\eeq
If we assume $\theta=\pi/2N$,
the photon that comes from the $N$th beam splitter goes to
the upper right port of $a$ with probability $1$.

Next, we consider the case where there is an object that absorbs
the photon on the paths $a$.
We assume that the object is put on every path $a$ that comes
from each beam splitter,
and all of these $N$ objects are the same one.
The photon thrown into the lower left port of $b$
cannot go to the upper right port of $a$
because the object absorbs it.
If the incident photon goes to the lower right port
of $b$, it has not passed through paths $a$ in the interferometer.

Therefore, the probability that the photon goes
to the lower right port of $b$ is equal to the product
of the beam splitters' reflectivities.
It is given by $P=\cos^{2N}\theta$.
In the large $N$ limit,
$P$ approaches $1$ as follows:
\beq
\lim_{N\rightarrow\infty}P
=\lim_{N\rightarrow\infty}\cos^{2N}(\frac{\pi}{2N})
=\lim_{N\rightarrow\infty}
[1-\frac{\pi^{2}}{4N}+O(\frac{1}{N^{2}})]
=1.
\eeq

From the above discussion,
we can conclude that Kwiat et al.'s interferometer
directs an incident photon from the lower left port of $b$
with probability $P$ at least as follows:
(1) if there is no absorptive object in the interferometer,
the photon goes to the upper right port of $a$;
(2) if there is the absorptive object in the interferometer,
the photon goes to the lower right port of $b$.
Furthermore, if we take large $N$, we can set $P$ arbitrarily
close to $1$.

\section{The Bell-basis measurement by the IFM process}
\lab{bell-basis-measurement-IFM}
Kwiat et al.'s IFM directs an outgoing photon
from the interferometer to the ports of $a$ and $b$
in Fig.~\ref{KWinterferometer3}
according to whether or not an absorptive object exists
on the paths $a$.
It admits the interpretation that information about the object
is written in the photon.
Moreover, because there is no annihilation of the photon
at the limit of $N\rightarrow\infty$
where $N$ is the number of beam splitters in the interferometer,
reduction of the state does not occur during Kwiat et al.'s IFM.
This means that there is no dissipation
and the system retains coherence
under the limit of $N\rightarrow\infty$.

From the above consideration,
we can regard the absorptive object as quantum
rather than classical.
In Sec.~\ref{introduction},
we consider the object to be a classical one
that can take only one of two cases:
the case where the object exists and the case where it does not exist.
In this section,
we treat the object as a quantum one and consider
that it can take a superposition of the two orthogonal states
where the object exists and does not exist.

From now on, we consider Kwiat et al.'s interferometer
to be a kind of quantum gate as shown in Fig.~\ref{ifmgate3}.
$a$ and $b$ in Fig.~\ref{ifmgate3} correspond to the paths
for the photon from the upper left and the lower left ports
in Fig.~\ref{KWinterferometer3} respectively.
$a'$ and $b'$ in Fig.~\ref{ifmgate3} correspond to the paths
for the photon from the upper right and the lower right ports
in Fig.~\ref{KWinterferometer3} respectively, as well.
The absorptive object comes from the path $x$
and goes to the path $x'$ in Fig.~\ref{ifmgate3}.
If we put the object into the path $x$
and put the photon into the path $a$ in Fig.~\ref{ifmgate3}
at the same time,
the dissipation occurs in the system.
Hence, we put a black rectangle on the port of $a$
in the gate drawn in Fig.~\ref{ifmgate3}.
From now on, we call the symbol of Fig.~\ref{ifmgate3}
the IFM gate.
Sometimes we call paths $x$ and $x'$ control parts
and $a$, $b$, $a'$, and $b'$ target parts
in the IFM gate.

\begin{figure}
\begin{center}
\includegraphics[scale=0.9]{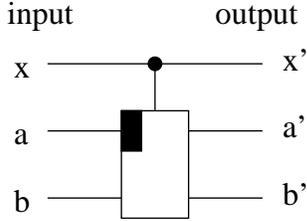}
\end{center}
\caption{The IFM gate that represents Kwiat et al.'s interferometer.}
\lab{ifmgate3}
\end{figure}

Table~\ref{Table-function-IFMgate} shows how the IFM gate
transforms states under the limit of $N\rightarrow\infty$.
The first line of Table~\ref{Table-function-IFMgate}
corresponds to a case where the absorptive object is not thrown
into the path $x$ and the photon is thrown into the path $b$.
The second line of Table~\ref{Table-function-IFMgate}
corresponds to a case where the absorptive object
and the photon are incident into the path $x$ and the path $b$
respectively.
We pay attention to the following fact.
When we do not put the object into the path $x$ and put the photon
into the path $a$,
the wave function is multiplied by a phase factor $(-1)$
as the third line of Table~\ref{Table-function-IFMgate}.
If we throw the particles into the paths $x$ and $a$
at the same time,
the object absorbs the photon and the dissipation occurs.
Thus the IFM gate is not unitary.
From now on, when we think about the IFM gate,
we take the limit of $N\rightarrow\infty$ and assume that
the transformation of Table~\ref{Table-function-IFMgate}
is valid.

\begin{table}
\begin{center}
\begin{tabular}{|c|c|}
\hline
input & output \\
\hline
$|0\ket_{x}|0\ket_{a}|1\ket_{b}$ & $|0\ket_{x}|1\ket_{a}|0\ket_{b}$ \\
$|1\ket_{x}|0\ket_{a}|1\ket_{b}$ & $|1\ket_{x}|0\ket_{a}|1\ket_{b}$ \\
$|0\ket_{x}|1\ket_{a}|0\ket_{b}$ & $-|0\ket_{x}|0\ket_{a}|1\ket_{b}$ \\
$|1\ket_{x}|0\ket_{a}|0\ket_{b}$ & $|1\ket_{x}|0\ket_{a}|0\ket_{b}$ \\
\hline
\end{tabular}
\end{center}
\caption{The function of the IFM gate
under the limit of $N\rightarrow\infty$.}
\lab{Table-function-IFMgate}
\end{table}

To put the following discussion concretely,
we replace the photon with an electron $e^{-}$
and replace the object with a positron $e^{+}$.
If the electron and the positron approach each
other closely enough,
we assume that they are annihilated and a photon
is created with probability $1$.
We consider this phenomenon to be the absorption
of the electron by the positron.
Because the electron and the positron
are quantum particles,
we have to regard them as wave packets
whose fluctuations are given by
$|\Delta\mbox{\boldmath $x$}||\Delta\mbox{\boldmath $p$}|
\sim\hbar/2$.
We suppose that $\Delta r$ is a characteristic length
of the interaction between the electron and the positron.
That is, the distance between them has to be less than
$\Delta r$ for their annihilation.
In our discussion, we assume
$|\Delta\mbox{\boldmath $x$}|\ll\Delta r$.
Therefore, we can consider the electron and the positron
to be point particles.
We can make beam splitters and mirrors for the electron and
the positron from plates that have suitable potential barriers.
We can adjust a reflectivity, a transmissivity,
and a phase shift of the beam splitter by its potential barrier.

In this paper, we construct the logical ket vectors
$\{|\bar{0}\ket,|\bar{1}\ket\}$ of a qubit from two paths
$a$ and $b$.
Hence, only one particle must always be on either of the two paths
as $|\bar{0}\ket=|0\ket_{a}|1\ket_{b}$
and $|\bar{1}\ket=|1\ket_{a}|0\ket_{b}$.
This method for constructing a qubit is called
the dual-rail representation \cite{Chuang-Yamamoto}.

The simplest usage of the IFM gate is the generation
of the Bell state as shown in Fig.~\ref{peBellcircuit4}.
$H$ is a beam splitter that works as follows:
\beq
H:
\left\{
\begin{array}{rrr}
|0\ket_{x}|1\ket_{y} & \rightarrow &
(1/\sqrt{2})(|0\ket_{x}|1\ket_{y}+|1\ket_{x}|0\ket_{y})\\
|1\ket_{x}|0\ket_{y} & \rightarrow &
(1/\sqrt{2})(|0\ket_{x}|1\ket_{y}-|1\ket_{x}|0\ket_{y})
\end{array}
\right..
\lab{Hadamard-transformation}
\eeq
We put
$|\bar{0}\ket_{+}|\bar{0}\ket_{-}
=|0\ket_{x}|1\ket_{y}|0\ket_{a}|1\ket_{b}$
in a quantum circuit shown in Fig.~\ref{peBellcircuit4}
as an initial state.
In this circuit,
the positron goes along the paths $x$, $y$
and the electron goes along the paths $a$, $b$.
From Eq.~(\ref{Hadamard-transformation})
and Table~\ref{Table-function-IFMgate},
a state in the circuit of Fig.~\ref{peBellcircuit4}
varies as follows:
\beqa
&& |\bar{0}\ket_{+}|\bar{0}\ket_{-}
=|0\ket_{x}|1\ket_{y}|0\ket_{a}|1\ket_{b} \non \\
& \stackrel{H}{\longrightarrow} &
(1/\sqrt{2})
(|0\ket_{x}|1\ket_{y}+|1\ket_{x}|0\ket_{y})|0\ket_{a}|1\ket_{b} \non \\
& \stackrel{\mbox{\scriptsize IFM}}{\longrightarrow} &
(1/\sqrt{2})
(|0\ket_{x}|1\ket_{y}|0\ket_{a}|1\ket_{b}
+|1\ket_{x}|0\ket_{y}|1\ket_{a}|0\ket_{b}) \non \\
&& =(1/\sqrt{2})
(|\bar{0}\ket_{+}|\bar{0}\ket_{-}+|\bar{1}\ket_{+}|\bar{1}\ket_{-})
=|\Phi^{+}\ket.
\lab{Bell-generation-transformation}
\eeqa
Hence, the quantum circuit shown in Fig.~\ref{peBellcircuit4}
generates the Bell state $|\Phi^{+}\ket$.

\begin{figure}
\begin{center}
\includegraphics[scale=0.9]{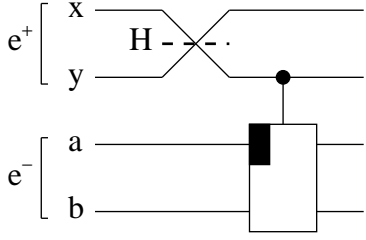}
\end{center}
\caption{A quantum circuit for generating the Bell state
of an electron and a positron.}
\lab{peBellcircuit4}
\end{figure}

The beam splitter $H$ in Fig.~\ref{peBellcircuit4}
applies the Hadamard transformation
to the qubit represented by the positron
on the paths $x$ and $y$.
Eq.~(\ref{Hadamard-transformation}) causes the following
transformation:
\beq
\left\{
\begin{array}{rrr}
|\bar{0}\ket_{+} & \rightarrow &
(1/\sqrt{2})(|\bar{0}\ket_{+}+|\bar{1}\ket_{+})\\
|\bar{1}\ket_{+} & \rightarrow &
(1/\sqrt{2})(|\bar{0}\ket_{+}-|\bar{1}\ket_{+})
\end{array}
\right..
\lab{Hadamard-transformation2}
\eeq
As shown above, by using a beam splitter,
we can realize an arbitrary $U(2)$ transformation
applied to one qubit that is given by the dual-rail
representation.

Then, let us consider the Bell-basis measurement
by the IFM gate.
Here, we consider the Bell states
made of the electron and the positron as
\beqa
|\Phi^{\pm}\ket
&=&
(1/\sqrt{2})
(|\bar{0}\ket_{+}|\bar{0}\ket_{-}
\pm|\bar{1}\ket_{+}|\bar{1}\ket_{-}), \non \\
|\Psi^{\pm}\ket
&=&
(1/\sqrt{2})
(|\bar{0}\ket_{+}|\bar{1}\ket_{-}
\pm|\bar{1}\ket_{+}|\bar{0}\ket_{-}),
\eeqa
where
$|\bar{0}\ket_{+}=|0\ket_{a}|1\ket_{b}$,
$|\bar{1}\ket_{+}=|1\ket_{a}|0\ket_{b}$,
$|\bar{0}\ket_{-}=|0\ket_{c}|1\ket_{d}$,
$|\bar{1}\ket_{-}=|1\ket_{c}|0\ket_{d}$,
the positron goes along the paths $a$, $b$,
and the electron goes along the paths $c$, $d$.
A quantum circuit shown in Fig.~\ref{IFMBellMeasurement3}
distinguishes these Bell states
$\{|\Phi^{\pm}\ket,|\Psi^{\pm}\ket\}$.

\begin{figure}
\begin{center}
\includegraphics[scale=0.9]{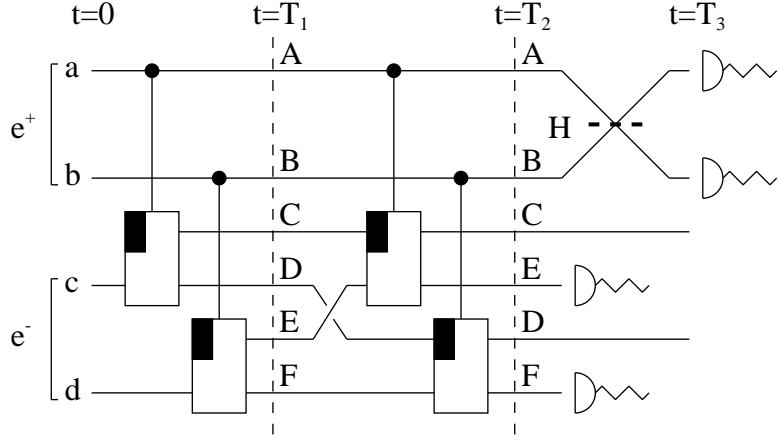}
\end{center}
\caption{A quantum circuit for the Bell-basis measurement.}
\lab{IFMBellMeasurement3}
\end{figure}

Let us consider how the quantum circuit
of Fig.~\ref{IFMBellMeasurement3} works.
We put one of $\{|\Phi^{\pm}\ket,|\Psi^{\pm}\ket\}$
into the paths $a$, $b$, $c$, and $d$ at $t=0$.
Table~\ref{Table-variation-BellMeasurement}
shows how the quantum circuit of Fig.~\ref{IFMBellMeasurement3}
transforms the basis vectors at $t=T_{1}$ and $t=T_{2}$.
The basis vectors are given by
$|\bar{0}\ket_{+}|\bar{0}\ket_{-}$,
$|\bar{0}\ket_{+}|\bar{1}\ket_{-}$,
$|\bar{1}\ket_{+}|\bar{0}\ket_{-}$,
and
$|\bar{1}\ket_{+}|\bar{1}\ket_{-}$
at $t=0$ as initial states.
Here we note that the path $D$ and the path $E$
are exchanged between $t=T_{1}$ and $t=T_{2}$.
We observe the electron on the paths $E$ and $F$
at $t=T_{2}$.
If we detect the electron on the path $E$,
we become aware that the initial state is $|\Psi^{\pm}\ket$.
On the other hand, if we detect the electron on the path $F$,
we become aware that the initial state is $|\Phi^{\pm}\ket$.
Because the electron is neither on the path $C$
nor on the path $D$
and the state is always $|0\ket_{C}|0\ket_{D}$,
we neglect these paths.

\begin{table}
\begin{center}
\begin{tabular}{|c|c|c|}
\hline
$t=0$ & $t=T_{1}$ & $t=T_{2}$ \\
\hline
$|0\ket_{a}|1\ket_{b}|0\ket_{c}|1\ket_{d}$ &
$|0\ket_{A}|1\ket_{B}|0\ket_{C}|0\ket_{D}|0\ket_{E}|1\ket_{F}$ &
$|0\ket_{A}|1\ket_{B}|0\ket_{C}|0\ket_{E}|0\ket_{D}|1\ket_{F}$ \\
$|0\ket_{a}|1\ket_{b}|1\ket_{c}|0\ket_{d}$ &
$|0\ket_{A}|1\ket_{B}|1\ket_{C}|0\ket_{D}|0\ket_{E}|0\ket_{F}$ &
$-|0\ket_{A}|1\ket_{B}|0\ket_{C}|1\ket_{E}|0\ket_{D}|0\ket_{F}$ \\
$|1\ket_{a}|0\ket_{b}|0\ket_{c}|1\ket_{d}$ &
$|1\ket_{A}|0\ket_{B}|0\ket_{C}|0\ket_{D}|1\ket_{E}|0\ket_{F}$ &
$|1\ket_{A}|0\ket_{B}|0\ket_{C}|1\ket_{E}|0\ket_{D}|0\ket_{F}$ \\
$|1\ket_{a}|0\ket_{b}|1\ket_{c}|0\ket_{d}$ &
$|1\ket_{A}|0\ket_{B}|0\ket_{C}|1\ket_{D}|0\ket_{E}|0\ket_{F}$ &
$-|1\ket_{A}|0\ket_{B}|0\ket_{C}|0\ket_{E}|0\ket_{D}|1\ket_{F}$ \\
\hline
\end{tabular}
\end{center}
\caption{Variation of states in Fig.~\ref{IFMBellMeasurement3}
at $t=T_{1}$ and $t=T_{2}$.}
\lab{Table-variation-BellMeasurement}
\end{table}

If we detect the electron on the path $E$,
the initial state $|\Psi^{\pm}\ket$ is projected
onto the following state at $t=T_{2}$:
\beq
(1/\sqrt{2})(-|0\ket_{A}|1\ket_{B}\pm|1\ket_{A}|0\ket_{B}).
\lab{projection-of-Psi}
\eeq
The function of the beam splitter $H$ is defined
by Eq.~(\ref{Hadamard-transformation}).
Hence, $H$ transforms Eq.~(\ref{projection-of-Psi})
as follows:
\beq
\stackrel{H}{\longrightarrow}
\left\{
\begin{array}{rr}
-|1\ket_{A}|0\ket_{B} & \mbox{for $|\Psi^{+}\ket$} \\
-|0\ket_{A}|1\ket_{B} & \mbox{for $|\Psi^{-}\ket$}
\end{array}
\right..
\eeq
Therefore, if we detect the positron on the path $A$
at $t=T_{3}$,
we find that the initial state is $|\Psi^{+}\ket$,
and if we detect the positron on the path $B$,
we find that the initial state is $|\Psi^{-}\ket$.

Next, we consider the case where the electron is detected
on the path $F$.
In this case, the initial state $|\Phi^{\pm}\ket$
is projected onto the following state at $t=T_{2}$:
\beq
(1/\sqrt{2})(|0\ket_{A}|1\ket_{B}\mp|1\ket_{A}|0\ket_{B}).
\lab{projection-of-Phi}
\eeq
The beam splitter $H$ transforms Eq.~(\ref{projection-of-Phi})
as follows:
\beq
\stackrel{H}{\longrightarrow}
\left\{
\begin{array}{rr}
|1\ket_{A}|0\ket_{B} & \mbox{for $|\Phi^{+}\ket$} \\
|0\ket_{A}|1\ket_{B} & \mbox{for $|\Phi^{-}\ket$}
\end{array}
\right..
\eeq
Therefore, if we detect the positron on the path $A$
at $t=T_{3}$,
we find that the initial state is $|\Phi^{+}\ket$,
and if we detect the positron on the path $B$,
we find that the initial state is $|\Phi^{-}\ket$.
From the above discussion,
we can distinguish the Bell basis vectors
$\{|\Phi^{\pm}\ket,|\Psi^{\pm}\ket\}$
by the IFM process.

\section{The CNOT gate operation by the IFM process}
\lab{cnot-gate-IFM}
Gottesman and Chuang showed that we can construct the CNOT gate
by preparing a state
\beq
|\chi\ket=(1/2)
[(|00\ket+|11\ket)|00\ket+(|01\ket+|10\ket)|11\ket],
\eeq
executing the Bell-basis measurement twice,
and carrying out one-qubit gate operations
according to results of the Bell-basis measurement
\cite{Gottesman-Chuang}.
Hence, we consider a method for generating $|\chi\ket$
here.

Fig.~\ref{IFM-4q-chistate3} shows a quantum circuit
for generating $|\chi\ket$.
We prepare
$|\bar{0}\ket_{+}|\bar{0}\ket_{-}|\bar{0}\ket_{+}|\bar{0}\ket_{-}$
as an initial state.
In Fig.~\ref{IFM-4q-chistate3},
the state varies as follows:
\beqa
&&
|\bar{0}\ket_{+}|\bar{0}\ket_{-}|\bar{0}\ket_{+}|\bar{0}\ket_{-} \non \\
& \stackrel{H}{\longrightarrow} &
(1/\sqrt{2})(|\bar{0}\ket_{+}+|\bar{1}\ket_{+})
|\bar{0}\ket_{-}|\bar{0}\ket_{+}|\bar{0}\ket_{-} \non \\
& \stackrel{\mbox{\scriptsize IFM1}}{\longrightarrow} &
(1/\sqrt{2})(|\bar{0}\ket_{+}|\bar{0}\ket_{-}
+|\bar{1}\ket_{+}|\bar{1}\ket_{-})
|\bar{0}\ket_{+}|\bar{0}\ket_{-} \non \\
& \stackrel{\mbox{\scriptsize IFM2}}{\longrightarrow} &
(1/\sqrt{2})(|\bar{0}\ket_{+}|\bar{0}\ket_{-}|\bar{0}\ket_{+}
+|\bar{1}\ket_{+}|\bar{1}\ket_{-}|\bar{1}\ket_{+})
|\bar{0}\ket_{-} \non \\
& \stackrel{H\otimes H\otimes H}{\longrightarrow} &
(1/2)[
(|\bar{0}\ket_{+}|\bar{0}\ket_{-}+|\bar{1}\ket_{+}|\bar{1}\ket_{-})
|\bar{0}\ket_{+}
+
(|\bar{0}\ket_{+}|\bar{1}\ket_{-}+|\bar{1}\ket_{+}|\bar{0}\ket_{-})
|\bar{1}\ket_{+}]
|\bar{0}\ket_{-} \non \\
& \stackrel{\mbox{\scriptsize IFM3}}{\longrightarrow} &
(1/2)[
(|\bar{0}\ket_{+}|\bar{0}\ket_{-}+|\bar{1}\ket_{+}|\bar{1}\ket_{-})
|\bar{0}\ket_{+}|\bar{0}\ket_{-}
+
(|\bar{0}\ket_{+}|\bar{1}\ket_{-}+|\bar{1}\ket_{+}|\bar{0}\ket_{-})
|\bar{1}\ket_{+}|\bar{1}\ket_{-}] \non \\
&&=|\chi\ket.
\eeqa
Here we use the following fact which can be derived
from Table~\ref{Table-function-IFMgate}.
The IFM gate causes a transformation as
$|\bar{0}\ket_{+}|\bar{0}\ket_{-}
\rightarrow
|\bar{0}\ket_{+}|\bar{0}\ket_{-}$
and
$|\bar{1}\ket_{+}|\bar{0}\ket_{-}
\rightarrow
|\bar{1}\ket_{+}|\bar{1}\ket_{-}$
when the positron is a control qubit and the electron is a target qubit.
As shown above, we can generate $|\chi\ket$.

\begin{figure}
\begin{center}
\includegraphics[scale=0.9]{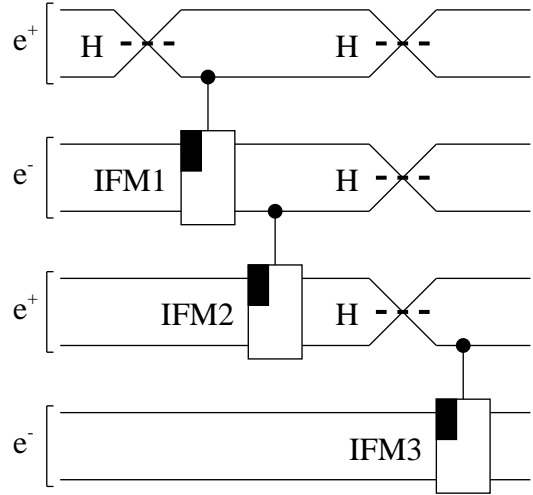}
\end{center}
\caption{A quantum circuit for generating $|\chi\ket$.}
\lab{IFM-4q-chistate3}
\end{figure}

Fig.~\ref{GCcnotcircuit3} shows a method for constructing the CNOT
gate proposed by Gottesman and Chuang.
A thin line transmits a qubit,
and a thick line transmits a classical bit.
(In Figs.~\ref{ifmgate3}, \ref{peBellcircuit4},
\ref{IFMBellMeasurement3}, and \ref{IFM-4q-chistate3},
a horizontal line of quantum circuits represents
a path along which a particle runs.)
$B_{1}$ and $B_{2}$ stand for the Bell-basis measurements.
An output of $B_{i}$ ($i\in\{1,2\}$) is given by
two classical bits:
$(x_{i},z_{i})=(0,0)$ for $|\Phi^{+}\ket$,
$(0,1)$ for $|\Phi^{-}\ket$,
$(1,0)$ for $|\Psi^{+}\ket$,
and $(1,1)$ for $|\Psi^{-}\ket$.
We apply $\sigma_{x}$ for $x_{i}=1$
and do nothing for $x_{i}=0$.
We apply $\sigma_{z}$ for $z_{i}=1$
and do nothing for $z_{i}=0$ as well.
We note that we can carry out any one-qubit unitary
transformation,
as $\sigma_{x}$ and $\sigma_{z}$,
by a beam splitter.

\begin{figure}
\begin{center}
\includegraphics[scale=0.9]{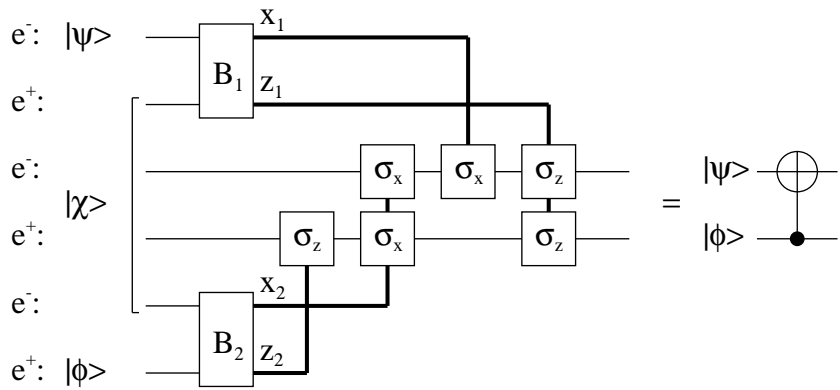}
\end{center}
\caption{Gottesman and Chuang's method for constructing the CNOT gate
by the Bell-basis measurement.
A thin line transmits a qubit,
and a thick line transmits a classical bit.}
\lab{GCcnotcircuit3}
\end{figure}

From the above discussion,
we find that we can apply the CNOT gate to an arbitrary state
of an electron $\forall|\psi\ket_{-}$
and an arbitrary state of a positron $\forall|\phi\ket_{+}$
using the IFM gates and beam splitters.
Then we have the following question.
How do we apply the CNOT gate to two arbitrary states of
electrons $\forall|\psi\ket_{-}$ and $\forall|\phi\ket_{-}$?
To solve this problem,
we use a technique shown in Fig.~\ref{pmswap2}.
(In Fig.~\ref{pmswap2}, a horizontal line represents a qubit
like Fig.~\ref{GCcnotcircuit3}.)
We apply the CNOT gate twice as shown in Fig.~\ref{pmswap2}
to the state of the electron $|\phi\ket_{-}$
and an auxiliary state of a positron $|\bar{0}\ket_{+}$.
Then, the wave functions are exchanged and we obtain
$|\bar{0}\ket_{-}$ and $|\phi\ket_{+}$.
After these operations, we can apply the CNOT gate to
$|\psi\ket_{-}$ and $|\phi\ket_{+}$.

\begin{figure}
\begin{center}
\includegraphics[scale=0.9]{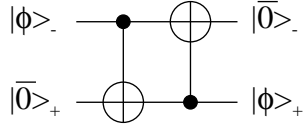}
\end{center}
\caption{A quantum circuit for exchanging wave functions of the electron
and the positron. A horizontal line represents a qubit.}
\lab{pmswap2}
\end{figure}

\section{Discussion}
\lab{discussion}
If we make a qubit by the dual-rail representation,
we can construct any one-qubit gate by a beam splitter.
In this paper, we show that we can construct the CNOT gate
by the IFM gates and beam splitters.
These facts imply that we can prepare a universal set of quantum gates
for quantum computation by the IFM.
Therefore, we can consider the IFM to be a basic operation
in the quantum information processing.

To carry out the Bell-basis measurement or to generate $|\chi\ket$
with probability $1$,
we need to take a limit of $N\rightarrow\infty$
where $N$ is the number of beam splitters in the interferometer
of the IFM gate.
In the limit of large $N$,
the transmissivity of the beam splitter
$T=\sin^{2}(\pi/2N)$ takes a very small value.
This implies that we need many of highly accurate beam splitters
to realize the IFM gate with high probability.

In this paper, we consider the IFM with an electron
and a positron.
To realize it, we need an accelerator.
We may construct the IFM gate by a conductive electron and
a hole in a semiconductor.

Horodecki investigated the dynamical evolution
of the IFM process, where the absorptive object is replaced
by an atom evolving coherently\cite{Horodecki}.
Plenio et al. discussed the generation of entanglement
between atoms inside an optical resonator
through the non-detection of photons.
This process resembles the IFM\cite{Plenio-etal}.

\bigskip
\noindent
{\bf \large Acknowledgements}
\smallskip

We thank M.~Okuda for encouragement.
We also thank M.B.~Plenio for drawing our attention
to Ref.~\cite{Plenio-etal}.

\end{document}